\documentclass[12pt]{iopart}
\thicklines
\input epsf.tex
\eqnobysec
\def\HA{H_{\rm{A}}}
\def\HB{H_{\rm{B}}}
\def\Hbig{H_{\rm{big}}}
\def\Hwee{H_{\rm{wee}}}

\def\qA{q_{\rm{A}}}
\def\qB{q_{\rm{B}}}
\def\qbig{q_{\rm{big}}}
\def\qwee{q_{\rm{wee}}}
\def\Nbig{N_{\rm{big}}}
\def\Nwee{N_{\rm{wee}}}
\def\cfour{c_{\rm{p+4}}}
\def\Nfour{N_{\rm{p+4}}}

\def\xperp{\bi{x}_\perp}
\def\parx{\bi{\partial}_\perp}
\def\parz{\bi{\partial}_z}
\def\para{\bi{\partial}_{\rm a}}
\def\parb{\bi{\partial}_{\rm b}}
\def\za{\bi{z}_{\rm a}}
\def\zb{\bi{z}_{\rm b}}
\def\zI{\bi{z}_{\rm I}}
\def\zz{\bi{z}}
\def\DD{\bi{\Delta}}
\def\dlz{\delta z}
\def\asxs{AdS$_5\times$S$^5$}
\def\rmd{\rm{d}}
\def\gs{g_{\rm{s}}}
\def\gYM{g_{\rm{YM}}}
\def\geff{g_{\rm{eff}}}
\def\ls{\ell_{\rm{s}}}
\def\be{\begin{equation}}
\def\ee{\end{equation}}
\def\ba{\begin{array}}
\def\ea{\end{array}}
\begin{document}
{\vbox{
\rightline{NSF-ITP-99-117}
\rightline{hep-th/9910098}}}
\title{Baldness/delocalization in intersecting brane systems}
\author{Amanda W. Peet}
\address{Institute for Theoretical Physics, University
of California, Santa Barbara, CA 93106-4030, U.S.A.}

\begin{abstract}
Marginally bound systems of two types of branes are considered, such
as the prototypical case of D$p\!+\!4$ branes and D$p$ branes.  As the
transverse separation between the two types of branes goes to zero,
different behaviour occurs in the supergravity solutions depending on
$p$; no-hair theorems result for $p\!\le\!1$ only.  Within the
framework of the AdS/CFT correspondence, these supergravity no-hair
results are understood as dual manifestations of the
Coleman-Mermin-Wagner theorem.  Furthermore, the rates of
delocalization for $p\!\le\!1$ are matched in a scaling analysis.

\bigskip
\noindent
Contribution to the proceedings of ``Strings '99''; based
on \cite{awpdm} written with D. Marolf.
\end{abstract}

\section{Introduction}

No-hair theorems appear prominently in the study of black holes.  They
are, however, not general; assumptions about the field content of the
action are required.

In the supergravity approximation to string theory, there are black
$p$-brane solutions available as well as black holes.  Certain systems
involving two types of BPS branes are marginally bound, stable at any
transverse separation.  Supergravity solutions which are fully
localized are known for some of these systems, albeit rarely in terms
of elementary functions.  In this article we will use the method of
Surya and Marolf \cite{SM} to exhibit the behaviour of the solutions
as the separation goes to zero, and to derive no-hair theorems.

The AdS/CFT correspondence \cite{Juan, IMSY} is a duality between
gauge theories on $p$-branes and the near-horizon limit of the
corresponding supergravity solutions.  We discuss this limit of the
supergravity solutions for the intersecting brane systems of interest,
and we find the dual phenomenon to the no-hair theorems to be the
Coleman-Mermin-Wagner theorem \cite{MW,col}, viz. the lack of
superselection sectors in low-dimensional quantum field theories.  We
also show that the delocalization rates obtained from a scaling
analysis in the field theory are captured by the supergravity; in
particular, the strong sensitivity to the dimension $p$.  We end with
remarks about future directions.

\section{Baldness, a.k.a. supergravity no-hair theorems}

One context in which a black hole no-hair theorem in General
Relativity may be seen is a process in which a small extremal black
hole of charge $q$ is brought up to the horizon of a large extremal
black hole of charge $Q\gg{q}$.  Since the geometry for the two
extremal black holes is static for any finite separation $\DD$, this
system can be easily studied at any $\DD$.  As the separation
$\DD\rightarrow{0}$, the configuration becomes spherically symmetric.
The theory does not allow the small charge to end up localized on the
horizon of the large black hole.

More generally, finding localized solutions in supergravity is
difficult.  As a very simple example, consider taking the D$p$-brane
supergravity solution and performing T-duality along a worldvolume
direction, say $x_{\rm p}$.  Use of the B{\"{u}}scher rules results in
a solution that looks like a D$p\!-\!1$-brane, except that it
possesses only a transverse SO($8\!-\!p$) symmetry instead of
SO($9\!-\!p$).  In other words, the harmonic function falls off like
$r^{p-7}$ instead of $r^{p-8}$.  This solution corresponds to the {\it
smeared} D$p\!-\!1$-brane.  Of course, we know what the unsmeared
D$p\!-\!1$-brane solution looks like in this case, but more generally
the unsmeared version of a smeared solution cannot be successfully
guessed.  The point of view which we advocate in this article is that
sometimes unsmeared supergravity solutions do not in fact exist, for
reasons that become clear in the context of the generalized AdS/CFT
correspondence.

In the following analysis we are interested in BPS systems comprising
two clumps of branes which are marginally bound to each other.  The
prototype of this system is the D0$\parallel\,$D4(0), and under
dualities this is connected to D$p\parallel\,$D$p\!+\!4$,
D$m\perp\,$D$n(p)$, $m\!+\!n\!=\!p\!+\!4$ e.g. D2$\perp\,$D2(0),
F1$\perp\,$D$p$(0), D$p\perp\,$NS5$(p\!-\!1)$, and so on.

Supergravity solutions for such systems have been studied by many
groups; a partial list is \cite{int0}-\cite{int11},\cite{ITY}, see
\cite{awpdm} for other references.  For purposes of orientation we
briefly review the features relevant for our discussion here.
Maintaining some generality for the moment, we let there be two types
of branes, D$A$ and D$B$, with common directions $t,\zI$, relatively
transverse worldvolume directions $\za,\zb$ respectively, and overall
transverse coordinates $\xperp$.  We use the ``harmonic function
rule'' \cite{int4} ansatz; suppressing the R-R gauge potentials this
is
\be\ba{ll}
{\rmd} s^2 =&\! \HA^{-1/2}\HB^{-1/2}(-{\rmd} t^2 + {\rmd}\zI^2) +
\HA^{1/2} \HB^{-1/2}({\rmd}\zb)^2 \nonumber\\\ms 
&\!+ \HB^{1/2} \HA^{-1/2}({\rmd}\za)^2 +
\HA^{1/2} \HB^{1/2}({\rmd}\xperp^2) \,; \\\bs
e^\Phi = &\! g_s \HA^{(3-A)/4} \HB^{(3-B)/4} \,.
\ea\ee
Allowing the branes to have harmonic functions that depend on all
coordinates transverse to them, i.e. $\HA=\HA(\xperp,\zb),
\HB=\HB(\xperp,\za)$, the equations of motion become (see, for example,
\cite{int5})
\be\ba{ll}
\left[ \parx^2 + \HB\parb^2 \right] \HA(\xperp,\zb) =&\! \qA\,\delta(A) 
\,; \nonumber\\\ms
\left[ \parx^2 + \HA\para^2 \right] \HB(\xperp,\za) =&\! \qB\,\delta(B) 
\,; \nonumber\\\ms
(\parb \HA)(\para \HB) = 0 \,.
\ea\ee
From the last equation, it is clear that one clump of branes, say the
A-branes, is necessarily delocalized along the worldvolume of the
other clump.  For the prototype D0$\parallel$D4 this is no restriction
because there are no $\zb$.  However, for an orthogonal intersection
such as D2$\perp$D2(0) this is a restriction on the generality of the
ansatz; we will remark on the case of orthogonal intersections in the
last section.  Relabeling the D$p+4$ brane the `big' brane and the
D$p$ the `wee' one, and $\za\equiv \zz$, our basic supergravity
equations become
\be\ba{l}\label{eoms}
\left[ \parx^2 \right] \Hbig(\xperp) = \qbig\,\delta({\rm big}) \,;
\nonumber\\\ms
\left[ \parx^2 + \Hbig(\xperp) \parz^2 \right] \Hwee(\xperp,\zz) 
= \qwee\,\delta({\rm wee}) \,. \ea\ee
The first equation is the usual one for the big branes.  The second
equation is more difficult to solve; here we will use the method
of Surya and Marolf \cite{SM}.

We can now pose the problem in which we are interested.  Consider the
setup of $\Nbig$ big branes with $\Nwee$ wee branes separated from
them in the transverse direction by $\DD$.

\hskip1.5truein{\epsfbox{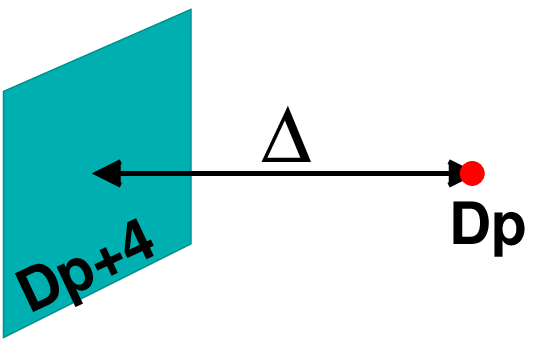}}  

\begin{center}
{\bf Figure 1}: the setup for our problem of interest. \\ The
D$p\!+\!4$-branes are displaced from the D$p$-branes by $\DD$.
\end{center}

\noindent
What we wish to know is whether the wee brane can be localized in the
worldvolume of the big brane as the separation is taken to zero, in
the supergravity regime.

To motivate this, consider the simple and familiar example of the
D-instanton in the background of the D3-brane.  The supergravity
solution is known explicitly only in the near-horizon region of the
D3-brane, which is \asxs, \cite{BGKR} (see also \cite{CHW,KL})
\be\ba{l}
{\rmd} S^2 = H_{-1} \left( {\rm AdS}_5\times{\rm S}^5 \right) \,;\quad 
\exp(\Phi) = \gs H_{-1} \,; \quad{\rm where} \nonumber\\\bs
{\displaystyle{ H_{-1} = 1 + {{3N_{-1}}\over{16\pi^6N_3^2\gs^2}} 
{{\rho^4\rho_0^4}\over{[{\rho_0}^2+|\bi{x}-\bi{x_0}|^2]^4}} }} \,.
\ea\ee
Here, the D-instanton position in Euclidean 10-space is
$(\bi{x_0},\bi{\rho_0})$, and the IR/UV relation relates $\rho$ to the
usual radial coordinate $U=r/\ls^2$ via
\be
\rho_0 = {{\sqrt{4\pi\gs N}}\over{U_0}} = 
{{\sqrt{4\pi\gs N\ls^4}}\over{\Delta}} = 
{{{\cal{R}}_{\rm AdS}^2}\over{\Delta}} \,.
\ee
The quantity $\rho_0$ is identified \cite{BGKR} as the scale size of
the instanton in the gauge theory.  Normally, the IR/UV relation is
interpreted as saying that the instanton scale size $\rho_0$ goes to
zero as the D-instanton goes to the boundary.  In the context of our
problem of interest, however, we are interested in the other end of
things: as $\Delta\rightarrow 0\,,$ $\rho_0\rightarrow\infty$.  An
alternative way to see that the D-instanton delocalizes without the
AdS/CFT baggage is simply to stare at the supergravity harmonic
function for the D-instanton: as the separation
$\Delta\propto{1/\rho_0}\rightarrow{0}$, $H_{-1}$ loses all
information about the relatively transverse coordinates $\bi{x}$
\be
H_{-1}
\quad\ba{c}{\mbox{\large{$\longrightarrow$}}}\cr\ns\ns
           {\mbox{\scriptsize{$\DD\rightarrow 0$}}}\ea\quad 1 \,.
\ee
This is a no-hair theorem as, if this $\bi{x}$ information had not
been erased, we could have used the above process to build black
D3-branes with D-instanton `hair' in the form of an arbitrary
distribution of D-instanton charge.

Note that in our analysis our solutions will always be taken to be
such that they can be matched onto the full asymptotically flat
solutions.  This means that in any $p$-brane AdS/CFT correspondence we
are doing gauge theory on $R^p$. As mentioned in \cite{awpdm}, this
condition rules out some known near-horizon supergravity solutions
\cite{Youm}; it would be interesting to learn what
interpretation these solutions may have in another context.  At this
point we also mention that there are some other known solutions which
are not of the type we want.  Some are partially localized, but
typically these are B{\"{u}}scher T-dual to larger-brane solutions.
In addition, from the second equation of motion in \eref{eoms} it is
possible to see that we can trade some transverse localization for
worldvolume localization.  Since this does not maintain the integrity
of the dimension of bigger brane in the solution, we will not consider
these either.

Let us recap the solution for the harmonic function of the big
(D$p+4$) branes.  We have
\be
\parx^2 \Hbig = \qbig\,\delta(\xperp) \quad\Rightarrow\quad
\Hbig = 1 + {{\cfour\gs\Nfour}\over{|\xperp|^{3-p}}} \,,
\ee
where $c_{\rm p}=(2\pi)^{(5-p)/2}\Gamma\left[(7-p)/2\right]$.  For the
wee branes,
\be
\left[ \parx^2 + \Hbig(\xperp)\parz^2 \right] \Hwee(\xperp, \zz) 
= \qwee\,\delta(\xperp-\DD)\,\delta(\zz) \,.
\ee
Surya and Marolf's method \cite{SM} for solving this equation of
motion involves doing a Fourier transform in the four relatively
transverse $\zz$ coordinates, and it allows transverse separation
$\DD$.  The Fourier transform converts the $\zz$ derivative into a
wavevector, and the equation can then be solved Fourier mode by
Fourier mode.  It can also be shown that in order to track the
behaviour of each Fourier mode at small separation $\DD$, keeping the
spherically symmetric modes suffices.  Then the equation reduces to a
second order ODE in the radial variable $r=|\xperp|$; this can be
solved explicitly and matched across the $\delta$-function.  The
resulting Fourier series/transform can be shown to converge
absolutely, except of course at the location of the $\delta$-function.
For details, the reader is referred to the original paper
\cite{awpdm}.

The results obtained by this method give distinctly different
behaviour depending on $p$ \cite{awpdm}:
\be\ba{cl}
-1\le p \le 1 :&\! {\rm baldness} \,; \nonumber\\
p=2 : &\! {\rm hair} \,. \nonumber
\ea\ee
(Recall that we are not going higher than $p=2$ because we take the
big-brane, which has dimension $p+4$, to be asymptotically flat.)  For
a previous idea on expectations based on the dimensionality $p$ see
\cite{Pelc}.

From the supergravity perspective these hair/baldness results come
about because of the different character of the differential equations
with $1/r$ versus stronger potentials $1/r^{n>1}$.  Note also that the
result for $p=2$ meshes nicely with the existence of an explicit
analytic solution found in \cite{ITY}, viz. D2-branes inside
D6-branes, in the near-horizon limit of the D6-brane.

We can actually do better than deriving no-hair theorems for the
low-dimensional cases; we can also work out the {\it rates} of
delocalization of the wee-branes as they approach the big-branes.
These may be computed by sitting out at a fixed transverse distance
from the big-brane, which we choose to be the characteristic
supergravity radius $R_{p+4}=\ls(\gs\Nbig)^{1/(3-p)}$, and watching
the behaviour of the different Fourier modes of the fields of the
wee-brane in the $\DD\rightarrow 0$ limit.  The delocalization can
thereby be characterized by a distance scale $\dlz$.  From the
supergravity equations we thusly find the scaling \cite{awpdm}
\be\ba{ll}
p=-1:&\! {\displaystyle{ 
     {{\dlz}\over{\ls}} \sim (\gs N_3)^{1/2} \,
     \left[ {{\ls}\over{\Delta}} \right] }} \,;
     \\\bs
p= 0:&\! {\displaystyle{ 
     {{\dlz}\over{\ls}} \sim (\gs N_4)^{1/2} \,
     \left[{{\ls}\over{\Delta}}\right]^{1/2} }} \,; \\\bs
p= 1:&\! {\displaystyle{ 
      {{\dlz}\over{\ls}} \sim (\gs N_5)^{1/2} \, \left\{
      \log\left[{{(\gs N_5)^{1/2}\ls}\over{\Delta}}\right]
      \right\}^{1/2} }} \,.
\ea\ee
The result for the D-instanton agrees with the IR/UV relation, which
is a consistency check.  The other relations look suspiciously
like low-dimensional quantum field theory relations, and we will
confirm this later in a precise fashion.  We will also see why the
no-hair theorems depend so strongly on the dimension $p$.  

\section{The AdS/CFT correspondence and the CMW theorem}

We would now like to define an AdS/CFT correspondence for this system
of two types of branes by analogy with \cite{Juan,IMSY}; see also
\cite{BPS}.  Of course, the gauge couplings on the big and wee branes
are built out of the same string theory parameters,
$\gYM^{2\,(p)}=(2\pi)^{p-2}\gs\ls^{p-3}$.  The decoupling limit
{\`{a}} la Maldacena can be cast in dimensionless statements as
follows: $(E\ls)\rightarrow{0}$, $E\sim{U=r/\ls^2}\sim\Delta/\ls^2$.
Now, since the dimensionless gauge coupling is formed by factoring in
$p\!-\!3$ powers of the typical gauge theory energy $E$ \cite{IMSY},
\be
\geff^2 = (2\pi)^{p-2}\gs (E\ls)^{p-3} \,,
\ee
the physics in the low-energy or decoupling limit will depend on which
dimensionless coupling we keep fixed.  Since we wish to keep the
physics of the big-brane in the game, we keep the big-brane coupling
fixed and so the wee-brane coupling will go to infinity in this limit,
at `fixed' energy $E$.  Simply put, the big-brane physics is
irrelevant in the infrared.  Thus, the field theory that we will be
interested in is $p\!+\!1$ dimensional.

We also need to pin down the other parameters.  Since we want to be in
the supergravity regime, we will keep $\Nwee, \Nbig \gg 1$ but
introduce no particular hierarchy between them.  We also choose to
keep the relative four-volume $V_4$ `fixed', i.e. we will not scale it
to zero as $\ls^4$.  We are most interested in the case where the
big-brane is non-compact, i.e. $V_4\rightarrow\infty$.  In the context
of the \asxs correspondence, this means having only a finite number of
D-instantons.  However, in our supergravity analysis we can also
handle the case of finite-volume, which in the \asxs case corresponds
to a finite density of D-instantons per unit D3-worldvolume; see
\cite{TL} for an analysis of this system.

\subsection{The supergravity side}

Taking the decoupling limit of our supergravity solution makes the big
($p\!+\!4$)-brane go near-horizon; we lose the 1 in its harmonic
function $\Hbig$,
\be
\Hbig\ \longrightarrow\
{{1}\over{\ls^4}}{{\cfour}\over{(2\pi)^{p+2}}}
{{\Nfour\,\gYM^{2\,(p+4)}}\over{U^{3-p}}} \,.
\ee
The more interesting issue is what happens to the wee-brane.  Our
solution retains dependence on the separation $\DD$.  For the cases
which delocalize, we can analyze the harmonic function for the
wee-brane near the core of the big-brane.  We find by careful
attention to boundary conditions that we must {\it not} drop the 1 in
the harmonic function for the delocalizing wee-brane \cite{awpdm}.
See also \cite{marsah} for similar conclusions on this issue.  This
phenomenon of not losing the 1 we saw very explicitly for the case of
the D-instanton in the near-horizon region of the D3-brane, again a
consistency check.  (For the $p=0$ system, in a sense we are operating
``half-way'' between a probe calculation where $H_0=1$ and a DLCQ
calculation where the 1 in the supergravity $H_0$ is dropped at the
outset.)

In addition, upon checking the dilaton and metric curvatures, we find
that the delocalizing wee-brane has a minor effect \cite{awpdm}; it
does not change the region where the curvature becomes large, and the
region where the dilaton becomes large occurs only extremely close to
the wee-branes - the presence of the big-brane transversely shrinks
the strong-dilaton region that occurs for $\Nbig=0$ .

Note also that for the case of finite $V_4$ it turns out that, as
$\DD\rightarrow 0$, we recover the smeared solution in which
$\Hwee\propto{1/U^{3-p}}$ just like $\Hbig$.

\subsection{The field theory side}

In the decoupling limit, the coupling between bulk and brane physics
vanishes.  In the $p\!+\!1$ dimensional field theory, decoupling
occurs between the Higgs and Coulomb branches, even quantum
mechanically \cite{Wit}.  For our situation where we are near-horizon
for the big-brane, we are in the Higgs branch of the field theory.
The field content may be recalled by remembering that the D$p$ branes
can be thought of as instantons in the D$p+4$-brane; the $p+1$
dimensional field theory is the theory of the collective modes of
these instantons.  As reviewed in \cite{hermanmicha} (see also
\cite{ABKSM},\cite{seiwit}), the fact that the wee-brane gauge 
coupling goes to infinity at fixed energy in our scaling limit means
that some fields, which correspond to the $p\!-\!p$ strings which are
transverse scalars, become auxiliary and may be integrated out.  The
resulting theory is a nonlinear sigma-model.  (See also
\cite{ofermicha} for an analysis in a different regime, where
different fields may be integrated out.)

The degrees of freedom of our Higgs branch theory are the instanton
scale size and orientation, and center of mass position.  In addition,
the theory is very strongly coupled because we are working in the
supergravity regime.  Let us concentrate on the instanton scale size
mode, which take as a proxy for how delocalized the wee-brane is.  In
low-dimensional quantum field theories, the Coleman-Mermin-Wagner
theorem prevents superselection sectors in the quantum regime.  This
theorem then tells us that we have wild infrared fluctuations leading
to complete delocalization for $p\le{1}$, i.e. baldness.  For
$p\ge{2}$ superselection sectors are allowed quantum mechanically, so
that we have hair for $p=2$.  These results are in direct accord with
the supergravity results, as we promised earlier.

In fact, we can go further and use scaling arguments to compute in the
low-dimensional field theories estimates for the delocalization rates.
As the separation $\Delta\rightarrow 0$, the IR fluctuations will
become stronger and stronger, and eventually give complete
delocalization.  Let us take the IR cutoff to be
\be
\Lambda_{\rm IR} = {{\Delta}\over{\ls^2}} \,.
\ee
In a sense we are reconstructing a field that was integrated out by
putting back this IR cutoff; see also \cite{hermanmicha} for a more
quantitative discussion on this point.

We now wish to calculate a rms value for the instanton scale size.  We
first treat the case of the D0-brane inside the D4-brane.  For a
single D0 instanton the moduli space metric is flat; we will assume
that for large-$N_0$ the free-field approximation works as well,
possibly modulo $1/N$ corrections which we will not worry about.  Also
we assume that each instanton fluctuates independently but has
approximately the same size - this latter assumption is motivated by
the work of \cite{doreyetal} on D-instantons in the original
AdS$_5$/CFT$_4$ correspondence.  Then, roughly speaking, since each
instanton has $N_4$ directions in the gauge group in which to point,
since the action has a normalization factor $\gs\ls$, and introducing
$\Lambda_{\rm IR}$ by dimensional analysis gives
\be\ba{cl}
\langle \rho^2 \rangle &\! \sim N_4 \ \gs\ls \ \Lambda_{\rm IR}^{-1} 
\,,\qquad {\rm i.e.} 
\nonumber\\\bs
{\displaystyle{ \left.
{\sqrt{\langle \rho^2 \rangle}\over{\ls}} \right|_{\rm{D0}} }}
&\! {\displaystyle{
 \sim (\gs N_4)^{1/2} \left[{{\ls}\over{\Delta}}\right]^{1/2} }} \,.
\ea\ee
This agrees with the supergravity result, as we are taking the rms
instanton scale size as a proxy for $\dlz$.  It also agrees with a
relation derived in \cite{hermanmicha}.  (We remark here that this
delocalization is not in contradiction with localization of center of
mass modes for wee-branes without big-branes present which occurs in
accordance with the existence of localized supergravity solutions for
wee-branes by themselves.  The difference in the current context is
partly in a factor of order $1/\Nwee\ll{1}$.  So even though it is
possible that the CM modes may be localized for the wee-brane
instantons inside the big-branes, the fact that the scale sizes are
totally delocalized is enough to conclude that we have agreement with
the supergravity results.)

For the $p=1$ case, by similar reasoning we get
\be
\langle \rho^2 \rangle \sim N_5 \ \gs\ls^2 \ 
\log\left[{{\Lambda_{\rm UV}}\over{\Lambda_{\rm IR}}}\right] \,.
\ee
Taking
\be
\Lambda_{\rm UV}\sim {{1}\over{\ls}} \,,
\ee
yields, to the accuracy of our scaling analysis,
\be
\left. {\sqrt{\langle \rho^2 \rangle}\over{\ls}} \right|_{\rm{D1}} 
\sim 
(\gs N_5)^{1/2} \left\{ 
\log\left[{{(\gs N_5)^{1/2}\ls}\over{\Delta}}\right] \right\}^{1/2}
\,,
\ee
which again agrees with supergravity.  In this case, renormalization
effects (which are absent for $p<1$) complicate the comparison with
\cite{hermanmicha}.

Although the case of the D-instanton is rather degenerate, in that the
field theory is zero-dimensional, the scaling analysis
\be
\langle \rho^2 \rangle \sim N_3 \ \gs \ \Lambda_{\rm IR}^{-2}  \,,
\ee
gives the scale-radius duality relation back again, up to constants:
\be
\left. {\sqrt{\langle \rho^2 \rangle}\over{\ls}} \right|_{\rm{D-1}}
\sim (\gs N_3)^{1/2} \left[ {{\ls}\over{\Delta}} \right] \,.
\ee
This is a consistency check on our method.

We have seen that the supergravity solutions for the parallel
intersections are capable of reproducing the field theory behaviour,
which is of course very sensitive to the dimension $p$.  Conversely,
the field theory analysis is capable of detecting supergravity no-hair
theorems.  

\section{Orthogonal intersections}

So far we have discussed the results of \cite{awpdm} only for the
prototypical cases D$p\,\parallel\,$D$p\!+\!4$.  We now discuss
briefly the case of orthogonal intersections, and for purposes of
illustration we discuss the example of D2$\,\perp\,$D2(0).

The field theory analysis of the previous section suggests that we can
draw hair/baldness conclusions for the near-horizon geometry of the
orthogonal intersections as well, by T-duality.  What matters is the
dimensionality $p$ of the intersection theory.  For example, for the
D2$\,\perp\,$D2(0) system which has a 0-dimensional intersection, we
expect baldness.  More specifically, the mode that used to be the
instanton scale size becomes, under T-duality, what might be termed
the `blow-up' mode.  If we trade the two worldvolume coordinates in
each clump of D2-branes (A and B) for a single complex coordinate $w$
so that we have $w_{\rm A},w_{\rm B}$, then the `blowup' mode $\rho$
appears in the combined worldvolume equation as $w_{\rm A} w_{\rm B} =
\rho^2$.

Now, in the supergravity regime, we have delocalization, but this
applies only to the near-horizon region of the geometry.  Let us
assume that there is no subtlety in matching this near-horizon part
onto the asymptotically flat part, as the supergravity equations of
motion suggest.  Then, we can draw conclusions about asymptotically
flat supergravity solutions as well.  Far away from the intersection,
we expect each clump of D2-branes to be localized, as they would be if
the other clump were not present.  Combined with smearing near the
intersection, this gives an overall picture of a supergravity geometry
with a `fat neck'.

More generally, for the solutions with intersection dimensionality
$p\ge{2}$, we expect to be able to have a skinny neck, but if
$p\le{1}$, we expect a fat neck.

The ansatz we made in setting up our supergravity analysis will not
cover the case of fully localized orthogonal intersections, because it
requires one clump of branes to be delocalized in the worldvolume of
the other clump.  Some progress has been made in generalizing the
ansatz, see for example \cite{int9}, but to our knowledge the only
explicit solutions obtained so far are perturbative \cite{donetal}.
In that analysis, the ansatz \cite{int9} was used, along with source
terms motivated by `BIon' physics of \cite{calda,gwg} and the
solutions were obtained to second order.  For $p\le{1}$ the
second-order perturbations were divergent, even after very careful
attention to boundary conditions and singularities, while for
$p\ge{2}$ they were well-behaved.  These results lend support to our
field theory expectations.  We await future progress in order to hear
the final verdict, e.g. to find out `how fat is fat'.

The search for various fully localized supergravity solutions is
difficult.  We would like to advocate the position that some of these
solutions may not be found because there is a good field theory reason
for them not to exist.

It would be interesting to see how far the results of our work can be
generalized.  One area of interest is nonextremal systems; work on
this is in progress.

\section*{Acknowledgments}

The author wishes to thank Don Marolf for collaboration on the work
presented at the conference.  In addition, we wish to thank Ofer
Aharony, Micha Berkooz, Per Kraus, Joseph Polchinski, and Herman
Verlinde for helpful discussions.

\bigskip\bigskip
\bigskip\bigskip

\end{document}